\documentclass[useAMS,onecolumn]{mn2e}

\usepackage{graphicx}
\usepackage{amsmath}
\usepackage{amssymb}

\newcommand{\EQ}[1]{Eq.(\ref{EQ #1})}  

\title{Early Afterglow  Emission from a Reverse Shock as a Diagnostic Tool for GRB Outflows}

\author[E. Nakar \& T. Piran]
  {Ehud Nakar,$^{1,2}$
  and Tsvi Piran,$^1$ \\
  $^1$Racah Institute for Physics, The Hebrew
University, Jerusalem 91904, Israel\\
  $^2$Institut d`Astrophysique de Paris, 75014 Paris,
France}

\begin{document}
\maketitle

\begin{abstract}
The Gamma-Ray burst (GRB) - afterglow transition is one of the
most interesting and least studied  GRB phases. During this phase
the relativistic ejecta begins interacting with the surrounding
matter. A strong short lived reverse shock propagates into the
ejecta (provided that it is baryonic) while the forward shock
begins to shape the surrounding matter into a Blandford-McKee
profile. We suggest a parametrization of the early afterglow light
curve and we calculate (analytically and numerically) the observed
parameters that results from a reverse shock emission (in an
interstellar medium [ISM] environment). We present a new
fingerprint of the reverse shock emission that is added to the
well known $t^{-2}$ optical decay. Observation of this signature
would indicate that the reverse shock dominates the emission
during the early afterglow. The existence of a reverse shock will
in turn imply that the relativistic ejecta contains a significant
baryonic component. This signature would also imply  that the
surrounding medium is an ISM. We further show that: (i) The
reverse shock optical flash depends strongly on initial conditions
of the relativistic ejecta. (ii) Previous calculations have
generally overestimated the strength of this optical flash. (iii)
If the reverse shock dominates the optical flash then detailed
observations of the early afterglow light curve would possibly
enable us to determine the initial physical conditions within the
relativistic ejecta and specifically to estimate its Lorentz
factor and its width.
\end{abstract}


\maketitle
\section{Introduction}

According to the internal-external shocks model (Piran \& Sari,
1998) the prompt gamma-ray burst (GRB) is produced by internal
shocks within a relativistic flow while the afterglow is produced
by external shocks between this flow and the surrounding matter.
The early afterglow appears during the transition from the prompt
$\gamma$-ray emission to the afterglow. During this transition the
relativistic flow, ejected  by the source, interacts directly with
the circum burst medium. This interaction can be used to pin down
the nature of the relativistic flow (baryonic or Poynting flux).
In a baryonic flow the reverse shock (RS) that propagates into the
ejecta produces both optical and radio emission. With Poynting
flux we expect only the higher energy forward shock (FS) emission.
While other sources of early optical and radio emission may exist
also in Poynting Flux flow, we show here that the RS emission has
a very robust optical and radio observable signatures that is very
unlikely to be imitated by other phenomena. If the flow is found
to be baryonic then the early afterglow signal could serve as a
diagnostic tool for the properties of the ejecta. This in turn,
would shed light on the nature of the inner engine that powers the
GRB.

Numerous authors (M\'esz\'aros \& Rees 1997; Sari \& Piran 1999a;
 Sari \& M\'esz\'aros  2000; Soderberg \& Ramirez-Ruiz 2003a;
Zhang, Kobayashi \& M\'esz\'aros 2003) considered the emission
from the reverse shock. Strong optical flashes in a rough
agreement with the RS predictions (Sari \& Piran 1999b;
M\'esz\'aros \& Rees 1999; Wang, Dai \& Lu 2000; Fan et al. 2002;
Soderberg \& Ramirez-Ruiz 2003b; Fox et al. 2003a; Weidong et al.
2003; Kumar \& Panaitescu 2003) were observed in two bursts (GRBs
990123 and 021211). In the first the RS predicted radio flare was
observed as well (Sari \& Piran 1999b; Kulkarni et al. 1999).  On
the other hand the early (the first $10$minutes) optical emission
observed in two other bursts (GRBs 021004 and 030418; Fox et al .
2003b; Rykoff et al. 2004) did not agree with the simple
predictions of an RS emission (see however Kobayashi \& Zhang
2003a). Furthermore, upper limits of $\sim 15$th mag  on the
prompt optical flux of several bursts (Williams et al. 1999;
Rykoff \& Smith, 2002; Klotz et al., 2003; Torii 2003a,b) have
lead to the so called ``optical flash problem": lack of bright
optical flashes (corresponding to RS emission) in many bursts.

Swift is expected to provide a large number of deep ($\sim$20 mag)
early ($\sim$minute) optical observations. We provide here
detailed predictions of the RS emission of a baryonic flow that
interacts with a constant density circum-burst medium (such as an
interstellar medium [ISM]). These predictions can be confronted
with the upcoming observations. We show that when the early
afterglow is found to be dominated by such RS emission (by passing
the observational tests) then the observations enable us to
determine the initial physical properties of the relativistic
outflow and to constrain the microscopic parameters in the
emitting region. We also show that the peak flux depends
sensitively on the strength of the reverse shock. It can vary over
more than five optical magnitudes between a mildly relativistic
and ultra relativistic shocks. Furthermore, previous calculations
have generally overestimated the peak flux of the mildly
relativistic RS by up to 7 optical magnitudes! In fact the FS
emission may even dominate over the RS emission at early times.
Altogether these results suggest a solution to the ``optical flash
problem". The calculated radio light curve shows that the radio
flare lasts long after the optical flash. We find that in the
generic case, in addition to the typical decay of the optical
flash (Sari \& Piran 1999a), the flash to flare time ratio and
intensity ratio provide another new test that the emission results
from a reverse shock.

Other mechanisms apart from the RS in an ISM environment can
produce an optical flash. Few examples are internal shocks
(M\'esz\'aros \& Rees 1997), a pair avalanche (Thompson \& Madau
2000, Beloborodov 2002),  RS  produced in a wind environment
(Chevalier \& Li 2000; Wu et al., 2003; Kobayashi \& Zhang 2003;
Kobayashi, M\'esz\'aros \& Zhang 2004) and even the FS as we
discuss here. While these mechanisms might be able to produce a
bright early optical flash, they are not expected to produce the
combination of the early optical and radio emission that we show
here to arise from an RS in an ISM environment \footnote{For
example, in the pair avalanche process the optical flash results
from a pair enriched FS and this emission is becoming harder with
time (Beloborodov 2002), in contrast to the RS emission. Thus, no
correlated radio flare is expected.}. Thus observing this
signature in many bursts in the future will solve further open
questions than just the flow's constitution (Baryonic vs. Poynting
flux). It will also reveal the circum-burst density profile and
determine the dominant mechanism that contributes to the optical
flash. In a separate paper (Nakar \& Piran, 04) we apply the tests
and the diagnostic tools presented here to GRB 990123. We find
that its early afterglow emission is remarkably consistent with
our predictions of the RS emission, suggesting strongly that at
least is this case the early afterglow is produced by a baryonic
flow propagating into an ISM.

The structure of the paper is somewhat unusual. The RS emission in
the most general case presents a large and complex diversity, but
the generic behavior is rather simple. Therefore, we begin in \S
\ref{sec:Generic} with a summary of the generic optical and radio
observables, and demonstrate how to use them in order to confirm
that the emission results from an RS and in this case to determine
the physical properties of the relativistic outflow. Later  we
describe the general analytical theory (\S \ref{Theory}) and the
numerical simulations (\S \ref{sec numeric}).

\section{The Generic Early Afterglow Light Curve}
\label{sec:Generic}

Consider a homogenous\footnote{The limits of the results presented
in this section when the shell is inhomogeneous are discussed in
sec. \ref{sec numeric}.} cold baryonic shell expanding
relativistically into an homogenous cold inter-stellar medium
(ISM). The problem is well defined by the shell's (isotropic
equivalent) energy $E$, width $\Delta$, initial Lorentz factor
$\Gamma_o$ and the ISM density $n$. As the ejecta shovels the ISM,
a forward shock and a reverse shock  are produced. The nature of
the RS is determined by the dimensionless parameter $\xi\equiv
(l/\Delta)^{1/2}\Gamma_o^{-4/3}$ (Sari \& Piran 1995, hereafter
SP95), where $l \equiv (3E/(4\pi nm_pc^2))^{1/3}$. If $\xi \ll 1$
the RS is relativistic and most of the shell's bulk motion energy
is dissipated in a single shock crossing of the shell, which occur
at a radius $R_\Delta \approx l^{3/4}\Delta^{1/4}$ (SP95). For
$\xi \gg 1$ the RS is Newtonian and many crossings are required to
dissipate a significant fraction of the energy. In the generic
case $\xi \lesssim 1$. In this case after the RS crosses the
ejecta once, the circum-burst gas, shocked by the FS, forms a
Blandford-Mckee profile and the original ejecta expands and cools
down at the tail of this profile (Kobayashi \& Sari 2000;
hereafter KS00). Therefore the hydrodynamical (and the radiative)
evolution is separated to two phases - during the RS and after the
RS. The emission from the separation point reaches the observer
simultaneously with a photon emitted on the line of sight from the
rear end of the shell at $R_\Delta$:
\begin{equation}\label{EQ t0}
    t_0=\left( \frac{\Delta}{c}+ \frac{R_\Delta}{2c\Gamma_o^2}\right) (1+z)=
    \frac{\Delta}{c}(1+0.5{\cal N}_t\xi^{3/2})(1+z) ,
\end{equation}
where $c$ is the light speed, $z$ is the redshift and  ${\cal N}_t
= 1.4$. Here and through out the paper ${\cal N}_x$ stands for
numerical correction factors to the analytic estimates (see sec.
\ref{sec numeric}).

The evolution before $t_0$ is highly sensitive to the initial
profile of the shell and in particular to the value of $\xi$. Thus
the observables before and at $t_0$ depend strongly on the initial
properties of the shell, and as such can be used as a diagnostic
tool of these properties. On the other hand, the RS that crosses
the shell erases, to a large extend, the initial shell profile.
Moreover, the evolution during the expanding and cooling phase
depends only weakly  on $\xi$ (KS00). Therefore, the behavior
after $t_0$ is insensitive to the initial conditions and as such
it provides a very unique and easily identified signature of an RS
emission.

The  general behavior of a basic RS (i.e. with no complications as
refreshed shocks during the RS etc.) over a broad spectrum is
described in Table \ref{tbl1}. In this section we summarize only
the generic behavior of the optical and radio observables and
demonstrate how to determine the physical parameters, $\xi$,
$\Delta$, and $\Gamma_o$, from the optical light curve. This
should be used with care as a non-generic behavior is always a
possibility.

The RS optical flash peaks at $t_0$. Therefore, similarity to late
afterglow parametrization (Beuermann et al. 1999), we suggest to
parameterize the RS optical emission as:
\begin{equation}\label{EQ FnurParam}
    F_{\nu,opt}^r(t)=F_0^r \left( \frac {1}{2} \left(  \frac{t}{t_0} \right) ^{-s\alpha_1}
    +  \frac {1}{2} \left( \frac{t}{t_0} \right) ^{-s\alpha_2} \right)
    ^{-\frac{1}{s}} ,
\end{equation}
where $\alpha_1>0$ and $\alpha_2<0$ are the power-laws indices of
a broken power-law that peaks at \footnote{For $s \gtrsim 1$ the
peak of Eq. (\ref{EQ FnurParam}) is obtained at $\approx t_0$ and
its value is $\approx F_0^r$. Numerical simulations show that in
the case analyzed here indeed $s \gtrsim 1$ (see sec. \ref{sec
numeric}). Thus, $t_0$ and $F_0^r$ that are found according to the
best fit of Eq. (\ref{EQ FnurParam}) can be taken, for any
practical purpose, directly from the observations as the time and
the Flux of optical peak.} $\approx F_{\nu,opt}^r(t_0)=F_0^r$. $s$
determines the sharpness of the peak. The optical frequency,
$\nu_{opt}$ is expected to satisfy
$\nu_a^r,\nu_m^r<\nu_{opt}<\nu_c^r$ where $\nu_a^r$ ,$\nu_m^r$ and
$\nu_c^r$ are the self-absorbtion, synchrotron and cooling
frequencies in the RS respectively. In this case (Sari \& Piran,
1999a):
\begin{equation}\label{EQ alpha2}
    \alpha_2 \approx -2.
\end{equation}
The decay slope, $\alpha_2$, as a post $t_0$ observable, is very
robust. KS00 show numerically that $\alpha_2 \approx -2$ for
various values of $\xi$ and $p$ and thus it is a signature of a
generic RS emission. On the other hand, $\alpha_1$ is most
sensitive to $\xi$. When $\xi \ll 1$, $\alpha_1 \approx 0.5$ and
as $\xi$ increases so does $\alpha_1$. For $0.05<\xi<5$ it can be
well approximated as (see Fig. \ref{plottwo}a):
\begin{equation}\label{EQ alpha1}
    \alpha_1 \approx {\cal N}_{\alpha,1}(0.5+\frac{p}{2}(\xi-0.07\xi^2)) ,
\end{equation}
where $p$ is the power-law index of the electrons' energy
distribution, and ${\cal N}_{\alpha,1}=1.2$ . Thus, a measurement
of $\alpha_1$ can determine the value of $\xi$ (up to the
uncertainty in the value of $p$). Once $\xi$ is known, Eq.
(\ref{EQ t0}), is solved for $\Delta$. Having $\xi$ and $\Delta$
one can find $\Gamma_o$ using:
\begin{equation}\label{EQ eta}
    \Gamma_o=188\xi^{-3/4}\Delta_{12}^{-3/8}(E_{52}/n)^{1/8} ,
\end{equation}
where we denote by $Q_{x}$ the value of the quantity $Q$ in units
of $10^{x}$ (c.g.s). Note that when $\xi \ll 1$ both Eqs. (\ref{EQ
t0}) and (\ref{EQ alpha1}) are insensitive to $\xi$ and only a
lower limit of $\Gamma_o$ can be found. $\Gamma_o$ depends very
weakly on the ratio $E/n$. Finally we find numerically that the
sharpness parameter $s$ depends strongly on the initial profile of
the shell, but not on $\xi$. The larger the initial Lorentz factor
dispersion ($\delta \Gamma_o = \Gamma_{o,max} /\Gamma_{o,min} $)
is the smaller is $s$ (wider peak). A homogenous shell ($\delta
\Gamma_o = 1$) results in a very sharp peak, $s \approx 10$, while
mild dispersion of $\delta \Gamma_o = 2$ may be sufficient to
reduce $s$ to  $\approx 1$.

It is remarkable that these initial parameters can be determined
without using $F_0^r$, and thus with no dependance on the poorly
known internal parameters, $\epsilon_{e}$ and $\epsilon_B$. The
value of $F_0^r$ can be used to constrain these parameters:
\begin{equation}\label{EQ F0r}
    \begin{array}{c}
        F_0^r=0.6{\rm mJy}~{\cal N}_F(1+z)^{-\frac{4+p}{8}}1.5^{2.5-p}
    \left(\frac{3(p-2)}{p-1}\right)^{p-1} \\
      \times\;\;\;\epsilon_{e-1}^{p-1}\epsilon_{B-2}^{\frac{p+1}{4}}n^{\frac{p+2}{8}}E_{52}^{1+\frac{p}{8}}
    t_{0,2}^{-\frac{3p}{8}}D_{28}^{-2}A_{F,0}^r(\xi), \\
    \end{array}
\end{equation}
where the numerical correction factor is ${\cal N}_F \approx 1/5$
and all the parameters and notations are as in Table \ref{tbl1}.
The function $A_{F,0}^r(\xi)$ is approximated in the range of
$0.1<\xi<2.5$ by:
\begin{equation}\label{EQ AF0rApprox}
    A_{F,0}^r(\xi) \approx 180 \xi^{0.65} \left(6 \cdot 10^{-4}
    \xi^{-2.6}\right)^\frac{p-1}{2}.
\end{equation}
The exact value of $A_{F,0}^r$ is given in Eq. (\ref{EQ AF0r}),
and must be used when $\xi$ is outside of the range above. $F_0^r$
depends strongly on $\xi$ and it varies by ~2 orders of magnitude
within the range most relevant for GRBs ($0.1<\xi<3$). The
relativistic and the Newtonian approximations (Fig.
\ref{plotone}b) overestimate $F_0^r$. Specifically the commonly
used  Newtonian approximation overestimate $F_0^r(\xi=1)$ by a
factor of 200. The numerical correction factor, ${\cal N}_F$, adds
another factor of 5!

The radio emission continues to rise after $t_0$ and it peaks at a
later time, $t_*$,  when $\nu_{radio}=\nu_a^r$. This happens
during the``cooling" phase of the shocked shell material.
Therefore the radio behavior  both before and after $t_*$ is a
robust feature (i.e. does not depend on the initial conditions).
The light curve depends only on the relations between $\nu_m^r$,
$\nu_a^r$ and $\nu_{radio}$ and the only remaining influence of
the initial conditions is via the values of the break frequencies
at $t_0$. Over a wide range of $\xi$ values
 $\nu_m^r(t_0) <\nu_a^r(t_0) \approx 10^{12-13}$Hz\footnote{A
 different case than the generic one(i.e different relation
 between $\nu_m^r$, $\nu_a^r$ and $\nu_{radio}$)is more likely
 here than in the optical emission, specially when the RS is
 ultra-relativistic.}.
In this case the radio flux at $t>t_0$ can be also characterized
by the parametrization of Eq. (\ref{EQ FnurParam}) with (see Table
\ref{tbl1})\footnote{As long as $\nu_{radio}<\nu_m^r$, $\alpha_r
\approx 0.5$. Observing the transition time to $\alpha_{r,1}
\approx 1.25$ determines $\nu_m^r(t_0)$. The early behavior at
$t<t_0$  can be found from Table \ref{tbl1}.}:
\begin{eqnarray}
\label{EQ Fradio}
      \alpha_{r,1} &\approx& 1.25  \;\;\;\;\;\;\;\;\;\;\;\;\;\;\;\;\;\; \alpha_{r,2} \approx -2 \nonumber\\
      \frac{t_*}{t_0}&=&\frac{\nu_a^r(t_0)}{\nu_{radio}} \\
      \frac{F_*}{F_0}&=&\left(\frac{\nu_a^r(t_0)}{\nu_{opt}}\right)^{-\frac{p-1}{2}}
      \left(\frac{\nu_{radio}}{\nu_a^r(t_0)}\right)^{1.3}.   \nonumber
    \end{eqnarray}
Eq. (\ref{EQ Fradio}) predicts a relation between the optical and
radio emission of the RS. It provides both an estimate of
$\nu_a^r(t_0)$ and a test that the emission results from an RS:
\begin{equation}\label{EQ radioTest}
\frac{F_*}{F_0}\left(\frac{t_*}{t_0}\right)^{\frac{p-1}{2}+1.3}=\left(\frac{\nu_{opt}}{\nu_{radio}}\right)^\frac{p-1}{2}
\sim 1000.
\end{equation}
Note that this value can be larger or smaller by a factor of $\sim
3$ (for a given $p$), because of the uncertainty in the post-RS
dynamics (KS00). Together with the optical decay signature, Eqs.
(\ref{EQ Fradio},\ref{EQ radioTest}) provide a unique imprint of a
baryonic RS. The determination of $\nu_a^r(t_0)$ (and maybe even
$\nu_m^r(t_0)$) provide additional constraints on $\epsilon_e$ and
$\epsilon_B$.


The early afterglow behavior described above is relevant only when
the RS is produced by interaction with an ISM like densities or
lower ($n \lesssim 100 \rm cm^{-3})$). This condition is not
satisfied if the circum-burst medium is the wind of a massive
star. For any reasonable parameters of such a wind the external
density during the crossing of the RS is few orders of magnitude
larger than this of a typical ISM. This brings the cooling
frequency well below the optical bands (Chevalier \& Li 2000) and
possibly the self absorbtion frequency above the optical band
(Kobayashi et al. 2004). This changes of the frequencies sequence
changes also the resulting behavior of the optical light curve
(e.g. the value of $\alpha_2$) . Therefore observing an early
afterglow that shows the RS emission signature described above
implies also that the density of the external medium is an ISM
like.

\section{Theory}
\label{Theory}

\begin{figure}
\begin{center}
\includegraphics[width=80mm]{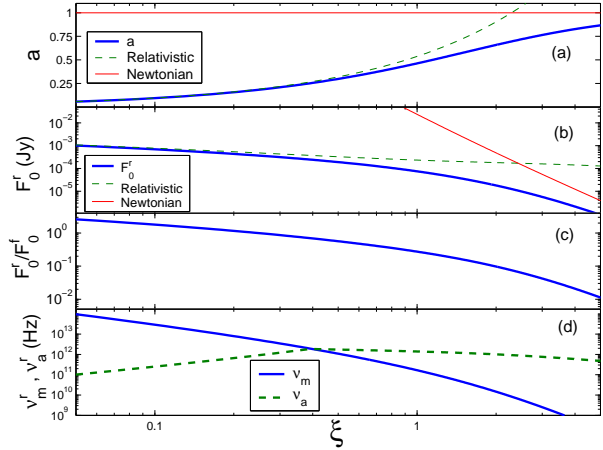}
\caption{\label{plotone} The parameter, $a$ and various break
frequencies and flux densities as functions of $\xi$. {\bf (a):}
The parameter $a$ according to Eq. (\ref{EQ a})({\it thick line}),
and its relativistic ({\it dashed line}) and  Newtonian ({\it thin
line}) approximations. {\bf (b):} $F_0^r$ according to Eq.
(\ref{EQ F0r} \& \ref{EQ AF0r}) when $a$ is calculated according
to Eq. (\ref{EQ a})({\it thick line}), and when the relativistic
({\it dashed line}) or Newtonian ({\it thin line}) approximation
of $a$ is taken. The physical parameters considered here are
similar to these considered in panel {\it d}. {\bf (c):} The ratio
between the RS peak flux and the FS flux at the same time.
$\epsilon_e$ and $\epsilon_B$ are assumed to be similar in the RS
and the FS. {\bf (d):} The synchrotron ({\it solid line}) and the
self absorbtion ({\it dashed line}) frequencies of the RS emission
at $t_0$. The parameters considered here are: $E_{52}=1$,
$\epsilon_e=0.1$, $\epsilon_B=0.01$, $n=1$, $p=2.5$, $t_0=100$sec
and $z=1$ ($D_{28}=1$).}
\end{center}
\end{figure}

The nature of the RS is determined by the dimensionless parameter
$\xi$, which in turn  determines the ratio $a$, of the Lorentz
factor of the shocked matter (in the explosion rest frame),
$\gamma_r$, to $\Gamma_o$:
\begin{equation}\label{EQ gammar}
    a \equiv \gamma_r / \Gamma_o .
\end{equation}
$a$ can be derived directly from the relativistic jump conditions
(SP95). It  satisfies\footnote{For a relativistic adiabatic
constant - $4/3$. Our results do not change significantly if  the
adiabatic constant varies smoothly between $4/3$ when $\xi \ll 1$
to  5/3 when $\xi \gg 1$}:
\begin{equation}\label{EQ a}
    (12/\xi^3-1)a^4+0.5a^3+a^2+0.5a-1=0.
\end{equation}
In the relativistic regime $a\approx a_R = 12^{-1/4}\xi^{3/4}$,
while in the Newtonian regime $a \approx a_N = 1$. Both
approximations overestimate $a$ in the intermediate regime and the
deficiency is largest when $\xi \approx 1$ (Fig.
(\ref{plotone}a)). For $\xi \lesssim$ a few the RS emission peaks
when the RS reaches the back of the shell. At this stage the
pressure, $p_r$, and the density, $n_r$, in the shocked shell as
measured in the shocked fluid rest frame, are:
\begin{equation}\label{EQ hydroRS}
     p_r=\frac{4}{3}a^2\Gamma_o^2 n m_pc^2 \\\ \\\ ;
    \\\ \\\ n_r=\xi^3n\Gamma_o^2 (2(a+1/a)/3+1).
\end{equation}
Assuming a  homogeneous initial shell and homogenous conditions
within the shocked region\footnote{These assumptions are relaxed
in the numerical simulations} these hydrodynamical conditions
determine $\nu_a^r$ ,$\nu_m^r$, $\nu_c^r$ and the peak flux
$F_{\nu,max}^r$ at $t=t_0$. These values appear in Table
\ref{tbl1} and can be used to estimate the radio and the optical
emission at $t_0$.  Using these values we derive the optical
emission, Eq. (\ref{EQ F0r}) and the exact value of
$A_{F,0}^r(\xi)$:
\begin{equation}\label{EQ AF0r}
    A_{F,0}^r(\xi)=770a^{4p-2}(a^2+1.5a+1)^{1-p}\xi^{3-3.75p}
    (1+0.5{\cal N}_t\xi^{\frac{3}{2}})^{\frac{3p}{8}}.
\end{equation}

\begin{figure}

\begin{center}
\includegraphics[width=80mm]{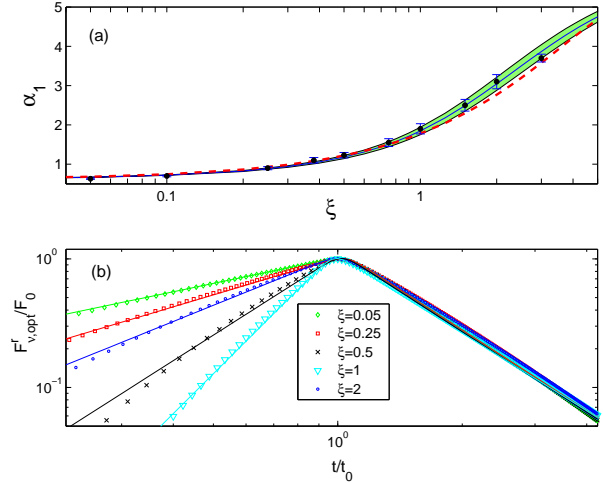}
\caption{\label{plottwo}Results of detailed hydro+synchrotron
numerical simulations. {\bf (a)}: $\alpha_1(\xi)$ as estimated by
the mean value of $dlog(F_{\nu,opt}^r)/dlog(t)$ at $t_0/2<t<t_0$,
according to Eqs (\ref{EQ tFrac},\ref{EQ FnuorFrac}) ({\it thin
line}), according to Eq. (\ref{EQ alpha1}) ({\it dashed line}) and
according to the numerical simulations ({\it dots}). The standard
deviation of $dlog(F_{\nu,opt}^r)/dlog(t)$ at $t_0/2<t<t_0$ is
depicted by the shaded area (Eqs \ref{EQ tFrac},\ref{EQ
FnuorFrac}) and the error bars (numerical simulations). {\bf (b)}:
Five numerical (normalized) optical light curves with various
$\xi$ values, and their best fits according to Eq (\ref{EQ
FnurParam}) (see       sec. \ref{sec numeric}).}
\end{center}
\end{figure}

Next we consider the evolution at $t<t_0$. The flux at $t<t_0$ can
be determined by parameterizing all the quantities according to
the fraction, $f$, of the shell that the RS has crossed:
$\Delta(f)$ and $E(f)$ $ \propto f$ while $n$ is constant. This
implies $\xi(f) \propto f^{-1/3}$ and $R(f) \propto f^{1/2}$. The
observer time
\begin{equation}\label{EQ tFrac}
    t(f) \propto f(1+0.5{\cal N}_t\xi(f)^{3/2})
\end{equation}
and the optical flux (for $\nu_a^r,\nu_m<\nu_{opt}<\nu_c^r$; see
Table \ref{tbl1}):
\begin{equation}\label{EQ FnuorFrac}
    F_{\nu,opt}^r(f) \propto f^pa^{4p-2}(a^2+1.5a+1)^{1-p},
\end{equation}
combine to yield $\alpha_1$. In the relativistic regime ($a
\propto \xi^{3/4} \propto f^{-1/4}$) $t \propto f$ and $F_0
\propto f^{1/2}$, hence $\alpha_1=0.5$. When $\xi$ increase the
logarithmic time derivative ($d\log(F_{\nu,opt}^r)/d\log(t)$)
varies with time, and its value at $t<t_0$ depends strongly on
$\xi$.  The description of the light curve as a power-law with an
index $\alpha_1$ is only an approximation. We estimate $\alpha_1$
as the mean value of $d\log(F_{\nu ,opt}^r)/d\log(t)$ during
$t_0/2<t<t_0$, and compare it to the standard deviation in this
time range (Fig. \ref{plottwo}a). The small deviation compared to
the mean value justifies the power-law approximation.

The evolution at $t>t_0$ is dictated by the post-RS hydrodynamics.
This hydrodynamics evolution was investigated both analytically
and numerically in KS00, and found to be almost independent of
$\xi$. We use the hydrodynamic evolution presented in KS00  to
determine the optical and radio light curves.

Finally we calculate the contribution of the forward shock
emission. The FS hydrodynamical conditions at $R_\Delta$ are:
\begin{equation}\label{EQ hydroFS}
    \gamma_f=\gamma_r \\\ \\\ ; \\\ \\\ p_f=p_r \\\ \\\ ;
    \\\ \\\ n_f=4a\Gamma_o n .
\end{equation}
The corresponding spectral parameters at $t_0$,
($\nu_{a,m,c}^f(t_0)$ and $F_{\nu ,max}^f(t_0)$), are listed in
Table \ref{tbl1}. The FS emission, in contrast to the RS emission,
depends rather weakly on $\xi$. Thus, the ratio between the two
vary strongly with $\xi$. Specifically,  the widely used relation
$\nu_m^f/\nu_m^r=(n_r/n_f)^2$ depends strongly on $\xi$. For $\xi
\approx 1 $ it indeed equals $ \approx \Gamma_o^2$. However, for
$\xi=0.1$ this  ratio equals $\approx 5 \cdot 10^{-4} \Gamma_o^2$.
Similarly, the ratio $F_{\nu,opt}^f(t_0)/F_{\nu,opt}^r(t_0)$
determines whether the RS or the FS dominates the emission. Fig.
\ref{plotone}c depicts this ratio for typical parameters. With
{\it these specific parameters}, and assuming similar $\epsilon_e$
and $\epsilon_B$ in both regions, the FS optical emission cannot
be neglected for $\xi> 0.5$.

\section{Numerical Simulations and Inhomogeneous Shells}
\label{sec numeric} The analytic calculations presented above
include several approximations.  In order to verify the accuracy
of the analytic calculations we carried out detailed numerical
simulations (Nakar \& Granot 2004) of the early afterglow
emission. We use these simulations to determine numerical
correction factors, denoted ${\cal N}_X$, for the analytic
calculations. The hydrodynamics simulations were done using a one
dimensional Lagrangian code that was provided to us generously by
Re'em Sari and Shiho Kobayashi (KS00).   The synchrotron radiation
code is described in Nakar \& Granot (2004). This code provides an
accurate synchrotron light-curve and spectrum, taking into account
the realistic hydrodynamical profile\footnote{We neglect the
feedback of the radiation energy loses on the hydrodynamics. This
is justified in the likely case that $\epsilon_e \ll 1$} of the
emitting region, the exact heating and cooling history of the
electrons and the precise photons arrival time to the observer
from each radiating element.

We have carried the simulations for a range of parameters with
$0.05 \leq \xi \leq 3$. Using these simulations we obtain
numerical corrections coefficients and  determine the accuracy of
the analytic estimates of $t_0$, $F_0^r$, $F_{\nu,opt}^f$ and
$\alpha_1$. We find that when including the numerical correction
factors (that range from 0.2 to 1.4), \EQ{t0} for $t_0$ is
accurate up to 10\% while \EQ{F0r} (for $F_{\nu,opt}^r$) and the
expression for $F_{\nu,opt}^f$ are accurate up to a factor $2$.
The sharpness parameter $s$ was not considered before, and we find
its value numerically. We find that if the shell is homogenous the
peak is very sharp, $s \approx 10$, regardless of the value of
$\xi$. Fig. \ref{plottwo}b depicts five numerical light curves
with $0.05\leq \xi\leq 2$, and their best fits according to
\EQ{FnurParam}. In all these fits $F_0$ is within factor of $2$ of
Eq.(\ref{EQ F0r}), $t_0$ is within 10\% of \EQ{t0}, $\alpha_1$ is
within the spread of \EQ{alpha1} (the shaded area in Fig.
\ref{plottwo}a), $\alpha_2 \approx -2$ and $s \approx 10$.

So far we have considered homogeneous shells. Clearly, the light
curve resulting from  an inhomogeneous shell would depend on the
shell's profile. To investigate partially the effect of
inhomogeneity, we have carried out numerical simulations of shells
with linear Lorentz factor profile ($\delta \Gamma_o=1.6$) and a
constant energy per a rest frame length interval. The value of
$\xi$ varied between $0.05$ and $1$ where $\Gamma_o$ is taken as
the mean value of the initial Lorentz factor. As expected in all
the cases we find $\alpha_2 \approx -2$ (which indicate that also
the radio emission should be insensitive to the shell's initial
profile) and a sharp decrease in $s$ when $\xi \approx 1$ ($s
\approx 1$ compared to $s \approx 10$ in the homogenous case).
$F_0^r$, $t_0$ and $\alpha_1$ are similar to the values obtained
in the homogenous case (a maximal difference of a factor of 1.5
between the homogeneous and inhomogeneous cases)\footnote{Note
that with a small value for $s$ (broad peak) $\alpha_1$ reaches
its ``asymptotic" value only far from the peak.}. These results
make us confidant that the solution we present here for homogenous
shell is generic and applicable also for inhomogeneous shells (at
least as long as $\delta \Gamma_o$ is not much greater than $1$,
and the Lorentz factor profile rises monotonically and regularly).

Finally we discuss briefly two caveats, which may arise naturally
in a baryonic relativistic flow, and may alter the RS emission and
produce a non-generic RS optical light curve. The first caveat is
a slow tail of the wind with a small but not negligible, amount of
energy (compared to the total energy of the wind). Such a tail may
results from the adiabatic cooling of the wind after the phase of
the prompt $\gamma$-rays emission, during which the wind must be
hot. In this case the peak may be obtained before the RS finishes
crossing the shell. Therefore the decay after the peak is expected
to be shallower then the generic value of $-2$ (even for several
orders of magnitude in time), becoming gradually steeper until the
generic value is obtained \footnote{ This may explain the shallow
early decay of GRB 021004. This scenario is very similar to the
continuous refreshed RS introduced by Sari \& M\'esz\'aros 2000,
and suggested by Fox et al. 2003b to explain GRB 021004. The only
difference is that here the slow tail of the flow is produced
naturally from the hydrodynamics of the relativistic wind and not
by a special source activity.}. The second caveat is a highly
irregular density profile. Such profile may result from a highly
irregular ejection process in the source, as expected in  the
internal shocks scenario when a burst is highly variable. Since a
hydrodynamic evolution smoothes the  pressure and the velocity
profiles, but not the  density profile, such irregularities are
expected to be carried by the flow also to the RS phase. If these
irregularities in the density profile are large enough, they are
expected to be reflected in the RS optical light curve during its
rising phase and before its decay reaches the asymptotical value.
Thus, a detailed observations of this phase may reveal the exact
profile of the ejecta, and maybe even be used to test the internal
shocks model (as the density profile in the flow is expected to
reflect the light curve produced by internal shocks). Practically,
if the early afterglow light curve is highly irregular, with no
underlying power-law, then the analysis method described here is
not applicable, and a theory that describes highly inhomogeneous
RS is required.  Note, however,  that if the asymptotic value of
the decay is reached soon after the peak, the tests of the RS
emission (Eqs. \ref{EQ alpha2} \& \ref{EQ radioTest}) are still
applicable.

\newcommand{\tb}[1]{{\small ${#1}$}}
\begin{table*}
\begin{minipage}{175mm}
 \caption{The Break Frequencies and Maximal Flux Densities}
  \label{tbl1}
 \begin{tabular}{lcllc}
 \hline
   & value at $t=t_0$ & $A_x^x(\xi)$ & at $t<t_0$ & $t>t_0$\\
 \hline
 \tb{\nu_m^r} & \tb{2\cdot
    10^{11}Hz (1+z)^{-\frac{1}{4}}\overline{\epsilon}_e^{\;2}\epsilon_{B-2}^\frac{1}{2}n^\frac{1}{4}
    E_{52}^\frac{1}{4}t_{0,2}^{-\frac{3}{4}}\overline{A}_{\nu,m}^{\;r}(\xi)}& \tb{a^8h_a^{-2}\xi^{-7.5}b_\xi^{3/4}} & \tb{\propto f^2a^8h_a^{-2}} & \tb{ -1.5} \\

    \tb{\nu_c^{r\;*}} & \tb{1\cdot 10^{17}Hz(1+z)^{-1/2}\epsilon_{B-2}^{-3/2}n^{-1}
    E_{52}^{-1/2}t_{0,2}^{-1/2}\overline{A}_{\nu,c}^{\;r}(\xi)} & \tb{a^{-4}\xi^{3}b_\xi^{-3/2}} & \tb{\propto f^{-2}a^{-4}b_\xi^{-2}}    \\

  \tb{\nu_a^{r\;\dagger}}& \tb{(5p+2)
  10^{12}Hz(1+z)^{-\frac{2}{5}}\overline{\epsilon}_e^{\;-1}\epsilon_{B-2}^{1/5}n^\frac{2}{5}
    E_{52}^\frac{2}{5}t_{0,2}^{-\frac{3}{5}}\overline{A}_{\nu,a}^{\;r}(\xi)} & \tb{a^{-8/5}h_a\,\xi^{12/5}
    b_\xi^{3/5}} & \tb{ \propto f^{-1}a^{-8/5}h_a} & \tb{-0.55}\\

    \tb{\nu_a^{r\;\dagger\dagger\dagger}}& \multicolumn{2}{l}{ {\footnotesize $(p-0.8)
    10^{13}Hz \left[(1+z)^{-\frac{p+6}{8}} \overline{\epsilon}_e^{\;p-1}\epsilon_{B-2}^\frac{p+2}{4}(n
    E_{52})^\frac{p+6}{8}t_{0,2}^{-\frac{3p+10}{8}} a^{4p}h_a^{\,1-p}\xi^\frac{6-15p}{4}b_\xi^\frac{3p+10}{8}\right]^\frac{2}{p+4}$}}
    & {\footnotesize $\propto \left[ \left(\frac{f}{h_a}\right)^{p-1}a^{4p}\right]^\frac{2}{p+4}$} & \tb{ -1}\\

  \tb{F_{\nu,b}^r} & \tb{250mJy(1+z)^{-\frac{5}{8}}\epsilon_{B-2}^{1/2}n^\frac{3}{8}E_{52}^{9/8}
    t_{0,2}^{-3/8}D_{28}^{-2}\overline{A}_{F,b}^{\;r}(\xi)} & \tb{a^2\xi^{-3/4}b_\xi^{3/8}} & \tb{\propto fa^2}
    & \tb{-0.95}\\ \hline

  \tb{\nu_m^f} & \tb{1\cdot 10^{16}Hz(1+z)^{1/2}\overline{\epsilon}_e^{\;2}\epsilon_{B-2}^{1/2}
    E_{52}^{1/2}t_{0,2}^{-3/2}\overline{A}_{\nu,m}^{\;f}(\xi)} & \tb{a^4\xi^{-3}b_\xi^{3/2}} &  \tb{\propto a^4} & \tb{ -1.5} \\

   \tb{ \nu_a^{f\;\dagger}}& \tb{ 3GHz(1+z)^{-1}\overline{\epsilon}_e^{\;-1} \epsilon_{B-2}^{1/5}n^{3/5}
    E_{52}^{1/5}\overline{A}_{\nu,a}^{\;f}(\xi)} & \tb{a^{2/5}\xi^{-3/10}}&\tb{\propto f^{3/10}a^{2/5}} & \tb{ 0}\\

  \tb{\nu_a^{f\;\dagger\dagger}} & \tb{0.6GHz(1+z)^{-\frac{1}{2}} \epsilon_{B-2}^{6/5}n^\frac{11}{10}
    E_{52}^{7/10}t_{0,2}^{-1/2}\overline{A}_{\nu,a}^{\;f}(\xi)} & \tb{a^\frac{22}{5}\xi^{-\frac{33}{10}}b_\xi^{\,\frac{3}{2}}}
    &  \tb {\propto f^\frac{13}{10}a^\frac{22}{5}b_\xi} & \tb{ -0.5}\\

    \tb{F_{\nu,b}^f} & \tb{1.5mJy(1+z)^{-1}\epsilon_{B-2}^{1/2}n^{1/2}E_{52}
    D_{28}^{-2}\overline{A}_{F,b}^{\;f}(\xi)}. & \tb{a^2\xi^{-3/2}} & \tb{\propto f^{3/2}a^2} & \tb{0} \\
\end{tabular}

\medskip
{\bf Notations -} ${\bf \epsilon_e,\,\epsilon_B}$: the fraction of
the internal energy in relativistic electrons and magnetic field
respectively; ${\bf p}$: the electrons spectral index; ${\bf D}$:
proper distance to the burst; $\overline{\epsilon}_e \equiv
30\epsilon_e(p-2)/(p-1)$; $b_\xi \equiv (1+{\cal
N}_t0.5\xi^{3/2})$; $h_a\equiv a^2+1.5a+1$; $\overline{A}_x^x(\xi)
\equiv A_x^x(\xi)/A_x^x(1)$

{\bf Using the table -} $F_\nu(\nu)$ is found according the
maximal flux, $F_{\nu,max}$, the values of the break frequencies
and the spectral power law indices between them. All these vary
between the different cases, where at each case $F_{\nu,max}$ is
at a different break frequency:
\newline
$\nu_m<\nu_a<\nu_c$ (generic case):
$F_{\nu,max}(\nu=\nu_a)=F_{\nu,b} (\nu_a/\nu_m)^{(1-p)/2}$; Power
law indices: $2<\nu_m<2.5<\nu_a<(1-p)/2<\nu_c<-p/2$
\newline
$\nu_a<\nu_m<\nu_c$: $F_{\nu,max}(\nu=\nu_m)=F_{\nu,b}$ ; Power
law indices: $2<\nu_a<1/3<\nu_m<(1-p)/2<\nu_c<-p/2$
\newline
$\nu_a<\nu_c<\nu_m$: $F_{\nu,max}(\nu=\nu_c)=F_{\nu,b}$ ; Power
law indices: $2<\nu_a<1/3<\nu_c<-1/2<\nu_m<-p/2$
\newline
{\bf Column (2)}: The values at $t=t_0$. {\bf Column (3)}: The
dependance at $t \leq t_0$ on $\xi$. {\bf Column (4)}: The
evolution at $t<t_0$ can be found using this column and Eq.
(\ref{EQ tFrac}) (see sec. \ref{Theory}). {\bf Column (5)}: The
approximated power-law indices at $t>t_0$, these values are
uncertain by at least $\pm 0.05-0.1$ (see KS00). The values are
calculated using the hydrodynamics of Eqs. (\ref{EQ hydroRS},
\ref{EQ hydroFS}) ($t<t_0$) and KS00 ($t>t_0$), and the radiation
calculations described in Sari, Piran \& Narayan (1998), Granot,
Piran \& Sari (1999) and Granot \& Sari (2001).
\newline
$^* \nu_c^f=\nu_c^r$ at $t<t_0\;\;\;\;\;\;\;\;\;^\dagger \;
\nu_a<\nu_m<\nu_c \;\;\;\;\;\;\;\;\; ^{\dagger\dagger}
\nu_a<\nu_c<\nu_m \;\;\;\;\;\;\;\;\; ^{\dagger\dagger\dagger}
\nu_m<\nu_a<\nu_c$

\end{minipage}
\end{table*}

\section{Conclusions}\label{sec conclusions}
We have suggested here a parametrization of the early optical
emission, as a broken power law with five parameters.  We have
calculated the values of these parameters for a RS produced by the
interaction of baryonic wind with a circum burst ISM. Our main
conclusions are: (i) The optical decay, ($\alpha_2 \approx -2$)
and the consistency between the peak time and flux of the optical
and the radio (Eq. \ref{EQ Fradio}) are robust features of a RS
emission in an ISM environment over a large range of initial
parameters. Observations of an early optical emission with these
features would suggest that: a) A significant fraction of outflow
energy is baryonic \footnote {Clearly a small fraction of the
outflow energy as a Poynting flux would not change the results
presented here, while a very large fraction of Poynting flux will.
The fraction of a Poynting flux energy needed to significantly
affect these results is yet unclear}. b) The circum-burst medium
is an ISM like. c) The RS emission is dominant over other possible
sources of optical flash. (ii) The values of the observables
before the optical peak depend strongly on the strength of the
shock, $\xi$, and can be used to pin down the initial conditions
of the flow. (iii) The combination of optical flash and radio
flare may also constrain the microscopic parameters in the
emitting region.

In addition to the specific optical predictions  we presented
detailed analytical results of the expected emission over the
whole spectrum. The advantages of these calculation over previous
ones are that they do not make any approximation on the strength
of the RS (i.e. relativistic or Newtonian), and that they are
confirmed (and corrected) by numerical simulations. The
conclusions of these calculations are: (i) Previous calculations
overestimated the intensity of the optical flash. Most pronounced
is the Newtonian approximation for $\xi \approx 1$ that
overestimate the optical flash by up to {\it three} orders of
magnitude. (ii)  An  optical flash brighter than $15$th mag is
expected in some but not in all GRBs. A GRB with typical
parameters and moderate energy ($E_{52}=1$), is expected to
produce a maximal flux of $R_{mag} \sim 17-19$ when the RS is
mildly relativistic ($\xi \approx 1$). (iii) Over some reasonable
range of the parameters space the FS emission dominates at all
times. (iv) When the RS is relativistic ($\xi \ll 1$) most of the
emission is released in the optical. When $\xi$ increase, the
emission is shifted to lower energy bands but it does not reaches
the radio, as $\nu_a$ in this case is $\sim 10^{12-13}$Hz. These
results suggest a solution to the ``lack" of optical flashes. Only
a fraction of the flashes is expected to be bright enough for
detection in the current observations, and in some cases an FS
emission is expected to dominate from the beginning.

We thank R. Sari and S. Kobayashi for providing us with their
relativistic hydrodynamics code and J. Granot, P. Kumar, R.
Mochkovitch, F. Daigne  and  E. Rossi for helpful discussions. The
research was supported by the US-Israel BSF and by EU-RTN: GRBs -
Enigma and a Tool. EN is supported by the Horowitz foundation and
by a Dan David Prize Scholarship 2003.

\end{document}